\documentstyle[12pt,aaspp4]{article}

\slugcomment{Submitted to {\em The Astrophysical Journal}}

\lefthead{Shahbaz et al.}
\righthead{T Pyxidis: The First Cataclysmic Variable with a Collimated Jet}

\newcommand{\gtsimeq}{\raisebox{-0.6ex}{$\,\stackrel
        {\raisebox{-.2ex}{$\textstyle >$}}{\sim}\,$}}

\begin{document}

\title{T Pyxidis: The First Cataclysmic Variable with a Collimated Jet}

\author{T.\ Shahbaz\altaffilmark{1}, M.\ Livio\altaffilmark{2},
K.\,A.\ Southwell\altaffilmark{1} and P.\,A.\ Charles\altaffilmark{1} }

\altaffiltext{1}{Dept.\ of Astrophysics, Nuclear Physics Building, 
Keble Road, Oxford, OX1~3RH, UK.\\ E-mail: tsh@astro.ox.ac.uk; 
kas@astro.ox.ac.uk; pac@astro.ox.ac.uk }
\altaffiltext{2}{Space Telescope Science Institute, 3700 San Martin Drive, 
Baltimore, MD 21218, USA.\\ E-mail: mlivio@stsci.edu} 

\begin{abstract}
We present the first observational evidence for a collimated jet in a
cataclysmic variable system; the recurrent nova T Pyxidis. Optical
spectra show bipolar components of H$\alpha$ with velocities $\sim$ 1400
km~s$^{-1}$, very similar to those observed in the supersoft X-ray
sources and in SS 433. We argue that a key ingredient of the formation of
jets in the supersoft X-ray sources and T Pyx (in addition to an
accretion disk threaded by a vertical magnetic field), is the presence of
nuclear burning on the surface of the white dwarf.

\end{abstract}


\section{Introduction}

\noindent
Highly collimated jets have been observed from a wide variety of
astrophysical objects. These include: young stellar objects (YSOs), high
mass X-ray binaries (HMXBs), black hole X-ray transients, possibly low-mass
x-ray binaries (LMXBs) containing a neutron star, symbiotic stars, planetary
nebula nuclei (PNN), supersoft X-ray sources (SSS) and of course active
galactic nuclei (AGNs) (see e.g. the recent review by Livio 1997 and
references therein).

An examination of the above list reveals immediately that all the systems in
which jets have been observed include accreting central objects. This has
led to the generally accepted paradigm, in which jets are formed (namely,
are accelerated and collimated) by accretion disks, which are threaded by a
reasonably ordered, large scale magnetic field (e.g.\ K\"onigl 1986; Blandford
1993; Pringle 1993; Livio 1997). Conspicuously absent in the above list, are
cataclysmic variables (CVs), even though they contain the best studied
accretion disks (see e.g. Warner 1995). In fact, winds and very 
poorly collimated
outflows have been observed from CV disks (e.g Drew 1995; Knigge, Woods \&
Drew 1996; Shlosman, Vitello \& Mauche 1996), but never a collimated jet. It
is unclear at present whether the fact that jets have not
been observed in CVs until now, merely represents an observational
difficulty (e.g. that there isn't enough material around CVs for the jets to
interact with, and hence become visible), or that CVs indeed cannot produce
jets.

In this letter we present the first observational evidence for a
collimated jet in the recurrent nova T Pyxidis. The observations and
results are presented in Sections 2 and 3, and their implications for jet
formation mechanisms are discussed in Section 4.

\section{Observations and data reduction}

We obtained intermediate resolution spectra of T Pyx and U Sco in 1996
and 1997 using the 3.9-m Anglo-Australian Telescope at Siding Spring,
Australia. The T Pyx spectra were taken in January 1996, 1997 and the U
Sco spectra were taken in June 1996. A Tek 1024$^{2}$ CCD attached to the
RGO spectrograph was used during all the observing runs. The spectral
resolution was 1.3\AA\ (FWHM=60 km s$^{-1}$ at H$\alpha$) for the 1200B/V
grating and 3\AA\ (FWHM=192 km s$^{-1}$ at H$\beta$) for the 600B
grating. Cu-Ar arc spectra were taken during the observations
at regular intervals for
calibration of the wavelength scale, in addition to a series of tungsten
flat-fields and bias frames. Further observational
details are given in Table 1.

The data reduction was performed using the Starlink {\sc figaro} package
and {\sc pamela} routines of K.\,Horne.  The bias level was removed from
each frame by using the mean overscan regions and then flat-fielded using
the observed tungsten lamp.
One-dimensional spectra were
extracted using the optimal algorithm of Horne (1986), 
and calibration of
the wavelength scale was achieved using the {\sc molly} package of
T.\,R.\ Marsh.

\placetable{tbl-1}

\section{Results and analysis}

In Fig.1 we show the variance-weighted average of the T Pyx spectra taken
during the two observing runs. The strong H$\alpha$ and He $\sc i$
emission lines are doubled-peaked and have broad bases (FWZI$\sim$ 1500
km~s$^{-1}$). We are also able to identify additional emission components
which we have marked as S$^{-}$ and S$^{+}$ (following the nomenclature
of Crampton et al. 1996). The wavelengths and relative velocities of the
S$^{+}$ and S$^{-}$ features as measured from the average spectra in Fig.
1 are listed in Table 2.

\placetable{tbl-2}

These spectral features have also been seen in the supersoft X-ray
sources (Crampton et al. 1996; Southwell et al. 1996), which are binary
systems in which a white dwarf accretes from a subgiant companion at such
a high rate that it burns hydrogen steadily at its surface (van den
Heuvel et al. 1992; Rappaport al. 1994) and in the prototype jet source
SS 433 (Margon 1984; Vermeulen 1993). 
By drawing the analogy of T Pyx with the
SSS we speculate that the observed bipolar features may be produced by
the same mechanism as in the SSS.

\placefigure{fig1}

\section{Discussion}

\subsection{The binary inclination of T Pyx}

\noindent
The inclination angle of T Pyx is not known. In the following we estimate
lower and upper limits to the inclination. Various authors have pointed
out that the double-peaked emission line profiles can be interpreted as
arising from an accretion disk viewed at high inclination. However, it
should be noted that a double-peaked emission line profile can also arise
from a system with an inclination as low as 15$^\circ$ (see the models of
Horne \& Marsh 1986). Assuming that the double-peaked lines arise
entirely from the disk, we can estimate the binary inclination $i$ of T
Pyx by measuring the separation of the H$\alpha$ emission-line peaks.

The Keplerian velocity of the outer edge of the disk ($v_{d}$) is given
by $v_{d}=(G M_{1}/R_{d})^{0.5}$. Combining this with Kepler's third law,
Paczynski's (1971) formula for the Roche lobe radius, and the assumption
that the accretion disk fills about 70 per cent of the white dwarf's
Roche-lobe, gives $v_{d}=1081 (M_{1}/P_{hr})^{1/3}$ km s$^{-1}$, where
$M_{1}$ is the mass of the white dwarf (in solar masses), and $P_{hr}$ is
the orbital period (in hours). The separations of the H$\alpha$
emission-line peaks measured from the average 1996 and 1997 spectra of T
Pyx are 202$\pm$6 km s$^{-1}$ and 180$\pm$17 km s$^{-1}$ respectively,
implying a projected velocity of the outer edge of the accretion disk of
$v_{d} \sin i \sim$ 100 km s$^{-1}$. Using the above formula with
$P_{hr}$=1.84 hrs (Schafer et al. 1992) and $M_{1} \sim 1.2 M_{\odot}$ (as
required for a recurrent nova, e.g. Webbink et al. 1987) gives a binary
inclination of $\sim 6^{\circ}$. This should clearly be regarded as a
lower limit to the binary inclination, since for very low inclinations 
thermal broadening in the disc may dominate the velocity distribution.

We can attempt to estimate the inclination angle also using the
observed outflow velocity (a similar procedure is often applied to YSOs,
e.g. Hirth, Mundt \& Solf 1994). Typically, outflows from CVs are observed
to have velocities in the range 3000-5000 km s$^{-1}$ 
(e.g. Drew 1995; these are
of the order of the escape velocity from the central object, e.g. Livio
1997). Taking the observed velocity of 1380 Km/s (Table 2), and an
average jet velocity of 4000 km s$^{-1}$, we obtain
i$\sim \cos^{-1}(Vobs/Vjet) \sim 70^{\circ}$.
This should be regarded as an upper limit since the system is
non-eclipsing.

\subsection{Should CVs have jets?}

\noindent
While it is not absolutely clear if the apparent absence of jets in CVs is
merely a consequence of observational difficulties (see Introduction),
suggestions have been made that jets cannot form in CVs (Spruit 1996).
This idea is based on the properties of one of the mechanisms suggested
for jet collimation; poloidal collimation. In this mechanism, the disk is
threaded by a vertical magnetic field which is strongest at the disk
center, but which has its largest flux in the outer disk.
The vertical outflow from the disk diverges at first, but as it
encounters the strong magnetic flux at the outer disk it is collimated (see
Blandford 1993, Spruit 1994, Ostriker 1997; Spruit et~al.\ 1997). Good collimation is obtained
if the disk radius is of the order of the Alfven radius (Konigl \& Kartje
1994; Spruit 1996; Ostriker 1997). For such a collimating configuration,
the minimum opening angle of the jet is given by: $\theta_{min} \sim
(R_{in}/R_{out})^{0.5}$, where $R_{in}$ and $R_{out}$ are the inner and
the outer disk radii respectively. An examination of the expected opening
angles for all the astrophysical objects which produce jets (Spruit 1996;
Livio 1997) reveals that whilst collimated jets are expected in AGNs,
YSOs, HMXBs, and black hole X-ray transients, they are not expected in
CVs, due to the relatively large ratio of $R_{in}/R_{out}$.
However, as was pointed out by Livio
(1997), the same argument suggests that jets would not have been expected
in SSS and in PNNs, and yet these two classes of objects do exhibit
collimated jets (Pakull 1994; Crampton et al. 1996; Southwell et al.
1996; Harrington \& Borkowski 1994; Trammell \& Goodrich 1996; Pollacco
\& Bell 1996). For example, the opening angle expected in the SSS RXJ
0513--69 is only smaller by a factor $\sim$1.5 than that expected for a
CV system with a 5 hr orbital period. The argument based on poloidal
collimation cannot therefore represent the complete picture.

Recently, it has been speculated that in order to produce {\it powerful}
jets, it is necessary to have, in addition to an accretion disk threaded
by a magnetic field, an energy/wind source associated with the central
object (Livio 1997). Livio then went on to identify this additional
energy source in every class of objects which are observed to produce
jets. In the case of the SSS and the PNNs, this additional source was
suggested to be nuclear burning on the surface of the white dwarf. It is
important to emphasize that the magnetized disk is still the key
ingredient for the acceleration and the collimation of the jet, with the
central source merely providing the extra bit of energy required to
ensure the existence of a solution with the desired properties (e.g.
Ogilvie 1997). According to this speculation, CVs would {\it normally}
not produce powerful jets since they have no such additional source (no nuclear
burning, no boundary layer, no supercritical accretion; see Livio 1997
for a discussion).

\subsection{The case for jets in T Pyx}

\noindent
Recurrent novae are cataclysmic variables observed to repeat nova
eruptions on a time scale of less than a century or so. T Pyx is unusual.
Its outbursts are very much like those of a slow nova (non
violent), whereas all other RNe have very fast nova outbursts. It has the
shortest orbital period (1.84 hr; Schaefer 1990) and in quiescence it has
extremely blue colours arising from a hot component. It is the nature of
the hot component which led Webbink et al. (1987) to suggest that nuclear
burning on the surface of the white dwarf was still taking place. In this
case the white dwarf would be very hot and so the thermonuclear runaway
would occur under not so degenerate conditions. This would lead to the
outburst being non-violent and slow, as is observed. A hot accretion disk
would not be sufficient to explain this property, (it should be noted
though that one could also explain the slow outburst in T Pyx if the mass
of the white dwarf was somewhat smaller than in the other RNe where it is
believed to be close to the Chandrasekhar limit; Webbink et al. 1987).
Therefore T Pyx may be the only CV in which nuclear burning on the
surface of the white dwarf is ongoing in quiescence. The detection of
jets from this system would therefore be consistent with Livio's (1997)
speculation, i.e. the comparison of nuclear burning RNe with the SSS.

Our optical spectra of T Pyx show features (marked S$^{+}$ and S$^{-}$ in
Fig. 1) which we interpret as high velocity components of H$\alpha$,
whose origin is in some type of bipolar outflow, similar to that observed
in the SSS. If our interpretation of these features and the analogy to
the SSS is correct, then T Pyx may be the first CV that shows collimated
jets. Because of the uncertainty in the inclination angle, it is
impossible at present to determine the degree of collimation. If we
assume that the entire width of the S features is due to finite
collimation (rather than to a spread in the velocity) and that the
outflow consists of a uniformly filled cone, then we obtain an opening
angle of $4.2^{\circ}$ for $i=70^{\circ}$, but very poor collimation for 
$i=6^{\circ}$. 
(The opening angle is approximately given by $\sin^{-1} \Delta V/(V \tan i)$, where
$\Delta V/V$ is the range in line-of-sight velocity of the S-features with
respect to the mean velocity, i.e. 20 percent; see Fig 1).

\placefigure{fig2}

\subsection{U Sco}

\noindent
An interesting test for the above scenario can be provided by another
recurrent nova, U Sco. According to our present understanding of
recurrent novae, the frequency of the outbursts can be understood in
terms of extremely high accretion rates ($ \gtsimeq 10^{-8}
M_{\odot}~yr^{-1}$) onto massive white dwarfs ($ \gtsimeq 1.2 M_{\odot}$;
see Webbink et al. 1987 and Prialnik \& Kovetz 1995). Therefore, the
accretion rates in T Pyx and U Sco should be comparable (since they have
a very similar outburst frequency). In U Sco, however, nuclear burning is
probably not taking place in quiescence, since the outbursts are very
violent, indicating a thermonuclear runaway under very degenerate
conditions (which would not have been the case if nuclear burning were to
continue in quiescence). Our spectroscopic observations (see section 2)
show that powerful jets are indeed not observed in U Sco (see Fig 2),
again in agreement with the need for an energy source associated with the
central object. However, this could be an observational selection effect,
since the system is eclipsing (Schaefer 1992), and therefore highly
Doppler-shifted components may not be observable.

\subsection{Comparing T Pyx with the supersoft sources}

\noindent
Although the observations show bipolar outflows from T Pyx
and the SSSs, and there is evidence for ongoing nuclear burning on the
surface of the white dwarf in both cases, the physical environment under
which this happens is different for the two objects. The mass accretion
rate in T Pyx is a factor of 10-100 less than in the SSS, this implies
that the nuclear burning is not steady (Nomoto 1982). Rather, T~Pyx
experiences nova outbursts, with episodes of continued burning after the
outbursts. This is not the case in the SSS, where the high mass accretion
rates are sufficient for the white dwarf to burn steadily. Another
similarity between T~Pyx and RXJ0513-69 is in the fact that there are
strong indications that the WD in the latter system is also fairly
massive (Southwell et~al.\ 1996).  

\acknowledgements

KAS acknowledges the support from PPARC. ML acknowledges support from
NASA Grants NAGW~2678 and GO-4377. We are grateful to the staff at the
AAT, Siding Spring. The data reduction was carried out on the Oxford
Starlink node, and the figures were plotted using the $\sc ark$ software.

{}

\begin{deluxetable}{llccc}
\tablecaption{Journal of observations. \label{tbl-1}}
\tablewidth{0pt}
\tablehead{
\colhead{Object} & \colhead{Date} & \colhead{Exposure time } & 
\colhead{Grating} & \colhead{Dispersion (\AA~pixel$^{-1}$)} }
\startdata
T Pyx   & 17/1/1996   	& 1$\times$900s  & 1200V  &  0.79 \nl
	& 5/1/1997  	& 10$\times$200s & 1200V  &  0.79 \nl
	& 5/1/1997   	& 5$\times$200s  & 1200V  &  0.79 \nl
U Sco   & 21/6/1996   	& 2$\times$1800s & 1200B  &  0.79 \nl
	& 21/6/1996   	& 3$\times$1200s & 600B   &  1.55 \nl
\enddata
\end{deluxetable}

\begin{deluxetable}{lcccc}
\tablecaption{ Properties of the H$\alpha$
Doppler-shifted components in T Pyx. \label{tbl-2}}
\tablewidth{0pt}
\tablehead{
\colhead{Year}	& \colhead{S$^{-}$ (\AA)} & \colhead{S$^{-}$ (km~s$^{-1}$)} 
&  \colhead{S$^{+}$ (\AA)} & \colhead{S$^{+}$ (km~s$^{-1}$)} } 
\startdata
1996	& -    & -      & 6593 &  1380 \nl
1997	& 6539 & -1082  & 6593 &  1380 \nl
\enddata
\end{deluxetable}

\clearpage

\figcaption[]{Average red spectrum of T Pyx taken in 1996 and 1997 with a
signal-to-noise ratio of $\sim$ 20 and 25 respectively. The principal
H$\alpha$ and He {\sc i} (6678 \AA) emission features are marked, along
with the Doppler-shifted components of H$\alpha$ labelled S$^{+}$ and
S$^{-}$. The spectra have been normalised to the continuum and have been
shifted vertically for clarity. The inset shows a close-up of the H$\alpha$
double-peaked emission line taken from the averaged 1997 spectrum.
\label{fig1}}

\figcaption[]{Average blue spectrum of U Sco with a signal-to-noise 
ratio of about 25 in the continuum. The He {\sc ii} (4541 \AA), 
He {\sc ii} (4686 \AA) and H$\beta$ (4861 \AA)
emission features are marked. No discernable Doppler-shifted components
can be seen. \label{fig2}}

\end{document}